# Role of current and daylight variations on small-pelagic fish aggregations around a coastal FAD from accurate acoustic tracking

M. Capello[1*], M. Soria[1], G. Potin[2], P. Cotel[1] and L. Dagorn[3]

1 UMR EME, Institut de Recherche pour le Développement (IRD), Saint Denis, La Réunion, France
2 ECOMAR, Laboratoire d'Ecologie Marine, Université de La Réunion, Saint Denis, La Réunion, France
3 UMR EME, Institut de Recherche pour le Développement (IRD), Victoria, Seychelles
∗ E-mail: manuela.capello@ird.fr

**Abstract**

We monitored twelve acoustically-tagged small pelagic fish (*Selar crumenophthalmus*) around a floating object in shallow water, playing the role of a coastal fish aggregating device (FAD). We characterized the response of the tagged-fish aggregation to varying current strengths and daylight. We found that the current induced a displacement of the aggregation upstream of the FAD, at distances that were increasing with the current strength. We gave evidence of an expansion and a higher coordination in the aggregation at dusk, with increasing swimming speed, distances among congeners and alignment. We discussed possible scenarios where fish polarization increases at dusk and proposed complementary measurements in future experiments that could confirm our findings.

## Introduction

Some pelagic fish species are known to associate with objects floating at the surface of the ocean (Hunter and Mitchell 1967). Although this phenomenon is largely exploited by fishers who use floating objects (usually called fish aggregating devices - FADs) to enhance their catch of various species (mainly tropical tunas), there is no scientific consensus on the origins of this behavior (Fréon and Dagorn 2000, Castro et al. 2002). With the increasing use of drifting FADs by industrial purse seiners all over the world, the scientific community has questioned the effects of FADs on the behavior of fish (Marsac et al. 2000, Hallier and Gaertner 2008). So far, most of the behavioral studies were conducted on tropical tunas (Dempster and Taquet 2004) and on their movements within a FAD array (Klimley and Holloway 1999, Ohta and Kakuma 2005, Dagorn et al. 2007). Studies on smaller species in shallow and coastal FADs (Soria et al. 2009, Capello et al. 2011), however, have recently revealed the potential of using fine-scale data (often difficult to obtain on larger species in the open ocean) to understand behavioral processes.

Fish aggregations could depend on several factors, such as the intrinsic characteristics of the object, oceanographic characteristics, abundance of prey, predators or conspecifics. The complexity of the problem is due to the probable high entanglement among all these factors and the difficulty of conducting accurate field-based measurements capable of quantifying them. Using fine-scale acoustic tagging data on a small pelagic fish around a single FAD, a recent study has quantified the main factors shaping the zone of aggregation (Capello et al. 2011). It turned out that the degree of sociability of individuals, as well as their swimming speed, strongly affect the radius of the aggregation. These findings came from observations in a 1-hour temporal window, during which the variability of the environmental conditions were negligible. Nevertheless, external factors, like current and daylight variation, could affect the behavior of fish when they are associated with a FAD. Although empirical observations done by fishermen assessed the importance of these quantities (in particular the effect of current on the position of fish in relation with the floating object), so far no studies have been able to quantify their effects on fish aggregations around FADs. In this paper, we investigate the effects of current and daylight variation on the behavior of a small pelagic fish species (*Selar crumenophthalmu*s) when associated with a floating structure.

## Material and Methods

### *Experimental setting*

Experiments were conducted in Saint Paul's bay, Reunion Island (South Western Indian Ocean). Forty bigeye scads (mean fork length 16.4 cm, SD 2.1 cm) were caught using hand lines, transported in baskets and maintained in tanks at the Aquarium of Reunion during ten days for acclimation. They were fed and treated with a solution of methylene blue and copper sulfate to kill bacteria and to prevent the proliferation of fungi. The tagging operation was carried out on the 1st of May 2003. We retrieved twelve individuals from the tank and we gastrically implanted acoustic tags by ingurgitation. The HTI$^{TM}$ acoustic tags (Model 795) were 7 mm diameter, 17 mm length and 1.5 g weight in the water. This weight was less than 0.1% of the mean fish weight and did not affect the buoyancy of the fish (Almeida et al. 2007). In situ tests led us to choose a pulse duration of 4 msec and the tag repetition rate was programmed to be different for each tagged fish and ranged between 1.43 and 1.16 s$^{-1}$.

The HTI$^{TM}$ Acoustic Tag Tracking System (Model 290) was composed of five hydrophones connected by cables to the Acoustic Tag Receivers system embedded on a boat of 12 m length fixed at 17 m depth by five anchors to prevent any movement. Four hydrophones surrounded the boat in a square of approximately 100 meters per side. The fifth hydrophone was located in the center of the square. The cables connecting the hydrophones to the boat reached the sea bottom straight below the boat, constituting a vertical submerged structure whose position was taken as our FAD position.

The twelve tagged fish were released on the 3rd of May 2003 at 12:00 in the proximity of the boat, anchored in Saint Paul's Bay, in the nearby of the fishing location, with 28 untagged congeners. The 3D tracking of each tagged fish (one position every second with sub-meter resolution) was possible within a radius of approximately 50 m from the FAD. All tagged fish stayed within the zone of detection until 19:00, with very few excursions outside the range of detection.

At the time the experiment was conducted, sunset was at 17:30. Current was measured through the Aanderaa RMC 9 Self Recording Current Meter, which was fixed under the boat (at 5 meters depth) in order to record the horizontal current speed and direction.

More details on the experimental setting can be found in (Capello et al. 2011).

*Methods for data processing and analysis*

Data processing involved two steps. First, the acoustic record of each tag on each of the five hydrophones was manually proofed using HTI$^{TM}$Mark Tags Software to exclude acoustic noise. Second, files were processed in HTI$^{TM}$Acoustic Tag program to track acoustic echoes and calculate fish positions through triangulation. The accuracy of the estimated position of fish in the horizontal plane ranged from 0.1 to 0.3 m in the monitoring network. In the vertical direction, the accuracy was lower (around 1.0 m).

Due to the system geometry, where the FAD was associated to a submerged vertical structure reaching the sea-bottom, the data analysis focused on the *xy* plane, neglecting the vertical direction. This approach was supported by previous acoustic survey measurements (Josse et al. 2000, Doray et al. 2006, Moreno et al. 2007) where the fish spatial distribution along z was not affected by the presence of the FAD but rather depended on the fish species.

Because we had detailed spatial information concerning several individuals, we calculated the pair-distribution function g(r) (Cavagna et al. 2008) among synchronous fish. We took synchronicity intervals of 1 second because this time frame was sufficient to obtain a large number of synchronous fish. We considered the time-averaged fish pair-correlation function g(r) at different hours of the day, which can be written as:

(Eq. 1) $$g(r) = \frac{1}{N(t)} \left\langle \frac{\sum_{ij} \delta(r - r_{ij}(t))}{2\pi r dr} \right\rangle,$$

where *N(t)* is the number of coplanar pairs detected in the temporal interval of one second [*t, t* + 1*s*] and $r_{ij}(t)$ is the planar distance among synchronous fish *i* and *j*. Coplanarity was established when two fish were within 1 m in the *z* direction, according to our experimental accuracy in the vertical direction. The delta function selects fish pairs at planar distance in the range [*r* − *dr, r*], with *dr* =0.3 *m*, and brackets denote the time average over one hour.

Another quantity that we calculated was the fish polarization in the *xy* plane:

(Eq. 2)
$$\phi = \frac{1}{N(t)} \left| \sum_i \frac{\vec{v}_i}{\|\vec{v}_i\|} \right|,$$

where the sum runs over the *N(t)* fish detected in the temporal interval [*t, t +* 1*s*], $\vec{v}_i$ is the planar speed vector of fish *i* and $\left|\vec{v}_i\right|$ denotes its modulus. This quantity is equal to 1 (0) when the group of fish is fully (un)polarized in the *xy* plane.

In the following, we denoted with hh:mm, the time interval of one hour between hh:mm and hh+1:mm (for example 13:00 denotes the time interval between 13:00 and 14:00).

**Results**

The fish trajectories in the *xy* plane showed clear differences according to the hour of the day, see Fig. 1.

In order to characterize this variability, we calculated the position of the tagged-fish center of mass and compared it with the average value of the current strength for each hour, see Fig. 2. We found that higher (lower) current strengths led to increasing (decreasing) distances between the tagged fish and the FAD (Fig. 2A), in a direction upstream of the FAD (Fig. 2B). We recorded a maximum in the current strength around 16:00 whereas the maximum displacement of the aggregated fish was found around 17:00. Equivalently, the increase of the current strength at 14:00 did not reflect a significant displacement of the tagged-fish center of mass, until 15:00. This suggests a possible delay of one hour in the fish response to the current.

Then, we characterized individual fish dynamics through the calculation of the swimming speed and the turning angle distribution averaged over all fish at the different hours (Fig. 3A and B). No temporal differences appeared in the turning angle, which manifested a robust peak around zero. Instead, the speed distribution showed temporal changes particularly marked at 18:00, with a shift in the speed distribution towards higher speeds.

For all times, g(r) showed a maximum around 0.6 +/-0.3 m, signaling that the most frequent fish-fish distance was always at short range (see inset of Fig. 4). Quite remarkable was the shape of the pair-distribution at 14:00, where the group of tagged fish showed a larger cohesion. Instead, at late hours, fish tended to occupy larger distances. This trend was smooth from 15:00 to 18:00, indicating a gradual expansion of the aggregation with time (Fig. 4).

Finally, we calculated the distribution of fish polarization at different hours. This quantity showed an enhancement of fish alignment at late hours, characterized by a smooth shift of the polarization towards higher values from 15:00 to 18:00, see Fig. 5.

**Discussion**

The above results clearly indicate that the fine-scale information obtained through the HTI acoustic technique is very promising for characterizing the role of different factors affecting the shape of small-pelagic fish aggregations around a FAD. Indeed, we could evaluate the effect of current variations on the position of the aggregation through the quantification of the displacement of the tagged-fish center of mass with respect to the location of the FAD. The effect of current on fish aggregation was clear, with the group moving upstream of the FAD at large distances when current increased. According to Lindquist and coauthors (Lindquist and Pietrafesa 1989), one possible reason for this orientation may be that the fish were choosing the location where they would likely find the least amount of current flow and therefore would have to spend the least amount of energy swimming in the current. This location could correspond to the upcurrent side of the FAD where incoming current flow is deflected back upon itself, resulting in narrow elliptical eddies that tend to negate current flow. A supporting argument, mentioned in (Lindquist and Pietrafesa 1989), is the advantage of facing into currents that bring food into the reef. Food that is missed on the first pass may be entrained in the upcurrent vortex reversal therefore making it possible for the bigeye scads to get several passes of the food organisms.

Additionally, we observed a delay of about one hour in the fish response to the current. However, this needs further investigations, since the maximum in the fish displacement observed at 17:00 (instead of 16:00 as found for the current) was not very pronounced. Equivalently, around 14:00, where a one-hour delay in the center of mass displacement was recorded, other anomalies were observed. Indeed, from the fish pair distribution, we found a contraction of the aggregate occurring at that time. This might signal that an unobserved event, like the presence of a predator or the arrival of another group of fish, occurred around 14:00, inducing a group contraction and a localization to the FAD much stronger than expected.

Similarly to what happens around 14:00, we could not assess the origin of the sudden increase in fish speed at 18:00, where possible unobserved predation effects might have affected the behavior of the tagged fish.

The effect of daylight emerged from the smooth variation of the tagged-fish pair-distribution and polarization observed when approaching dusk. Those changes could not be related to current effects, since around 15:00 and 17:00 the current strength and angle were nearly the same. From the pair correlation we observed an expansion of the aggregation at decreasing daylight. This fact, where fish aggregations are characterized by low packing densities at night, is well known from acoustic measurements (Fréon et al. 1996).

Additionally, we found that the polarization of the group of tagged fish was smoothly increasing towards dusk. On one side, this confirms that vision alone could not mediate fish alignment. Both lateral line and vision are employed simultaneously to maintain group polarization (Pitcher et al. 1976). On the other side, these results were quite surprising, since we expected the opposite scenario, with a gradual expansion in the fish aggregation inducing a lower polarization of the group. It is possible that other factors, not measured in this experiment, could be crucial in explaining these findings. For example, the number of congeners present in the aggregation could have increased from noon to dusk. Since we only quantified the tagged-fish behavior, we observed a consequent expansion in the aggregation. In this case, the fish aggregation would increase its size in the course of the day and manifest a smooth transition towards higher degrees of polarization at dusk, before the fish departure. Another possible explanation could be that group polarization depends on its distance to the FAD. In this scenario, the closer the group is to the FAD, the more the thigmotaxis to the FAD perturbs the alignment among congeners. In this way, the polarization of aggregated fish increases when the group moves away from the FAD, up to the fully-polarized limit of free schools.

If previous studies have shown that social effects are playing a fundamental role in the aggregations of small pelagic fish to FADs, we demonstrated that variable environmental conditions could strongly affect the properties of the aggregation. However, this study shows that, although changes in current and daylight can play a role, these factors alone cannot explain the full dynamics of the system observed during the day. Possible inter-specific interactions, predation effects, as well as an increasing number of congeners, might be important in determining the properties of the aggregation. Further investigations based on similar experimental settings, complemented with continuous visual inspections near the FAD, could confirm our picture of the aggregation dynamics.

**Acknowledgements**

We thank M. Taquet, L. Bigot, P. Durville, M. Timko, G. Fritsh for their help in fishing and tagging operations and P. Fréon for interesting discussions. This work was supported by the Regional Council of La Réunion Island, the French Ministry of Overseas, the 'Run Sea Science' (Theme Capacity Building) and the 'MADE: Mitigating adverse ecological impacts of open ocean fisheries' (Theme 2 – Food, Agriculture, Fisheries and Biotechnology, contract #210496) FP7 European projects.

**Figures and captions**

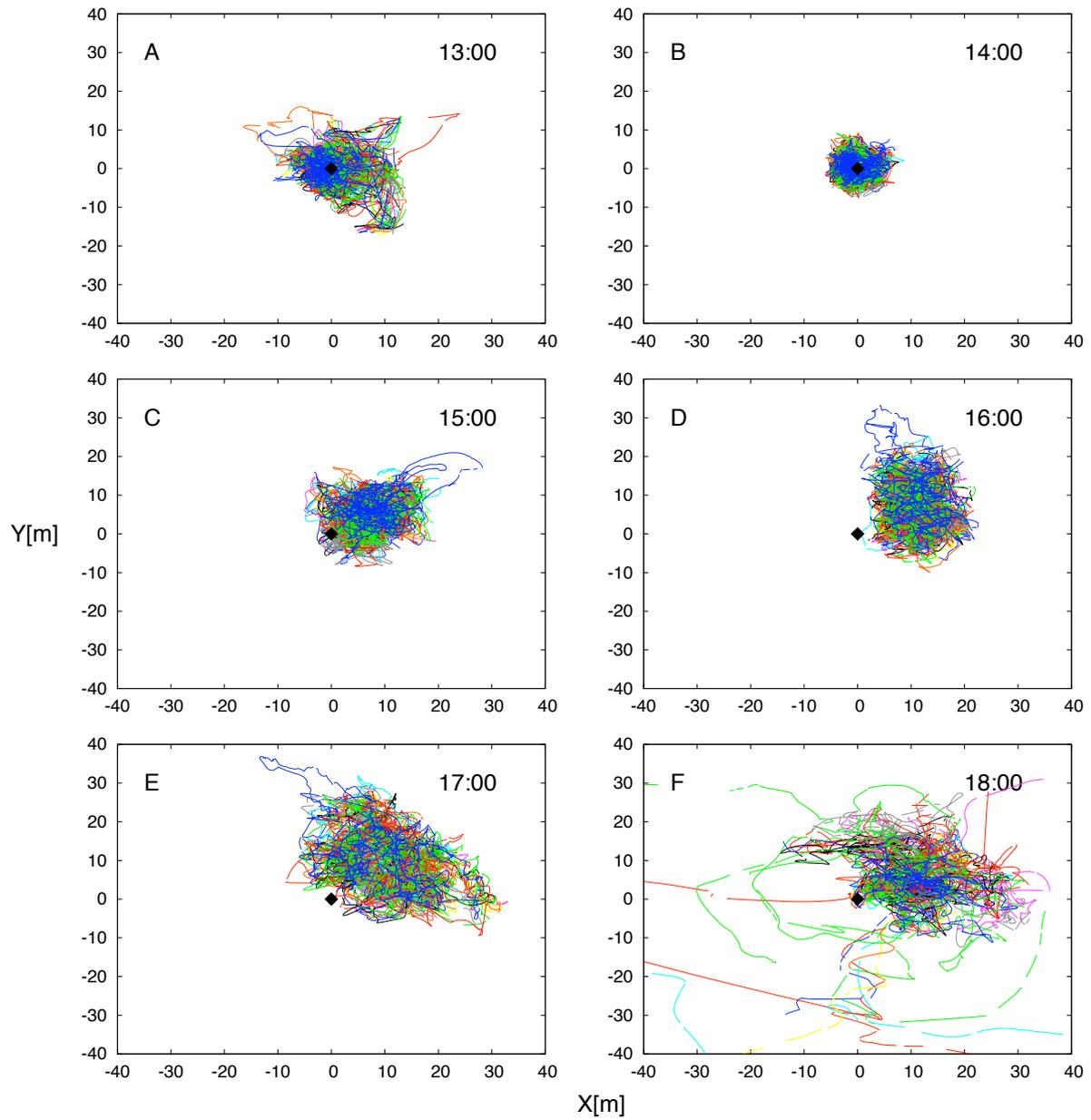

Figure 1: Fish trajectories in the xy plane at different hours of the day, from 13:00 (A) to 18:00 (F). Each color represents a different fish. The black diamond corresponds to the position of the FAD.

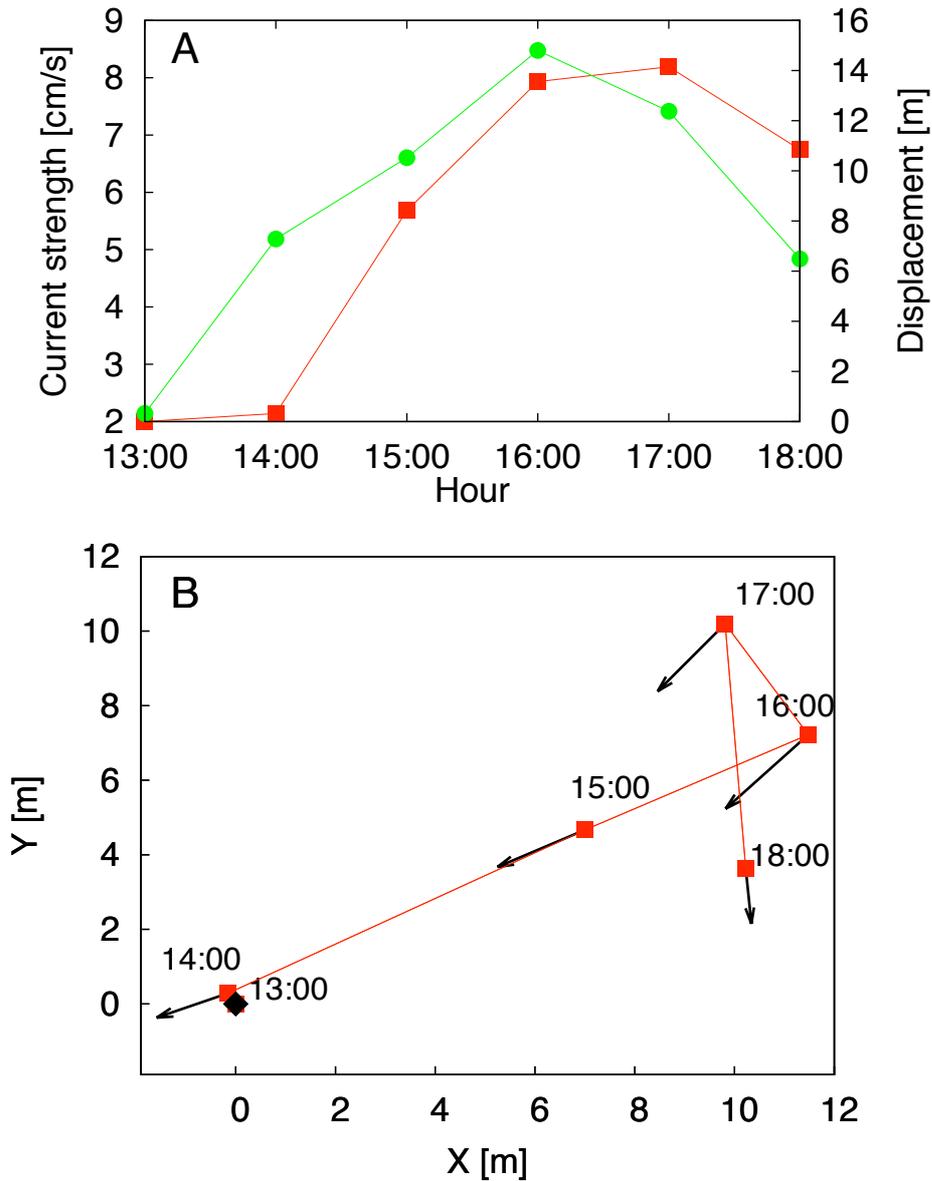

Figure 2: (A) Center of mass displacement (red squares, right axis) and average value of the current (green circles, left axis), calculated over time intervals of 1 hour, from 13:00 to 18:00. (B) Position of the fish center of mass in the xy plane at different hours of the day. Arrows indicate the current direction, with their length being proportional to the current strength. The black diamond indicates the FAD position.

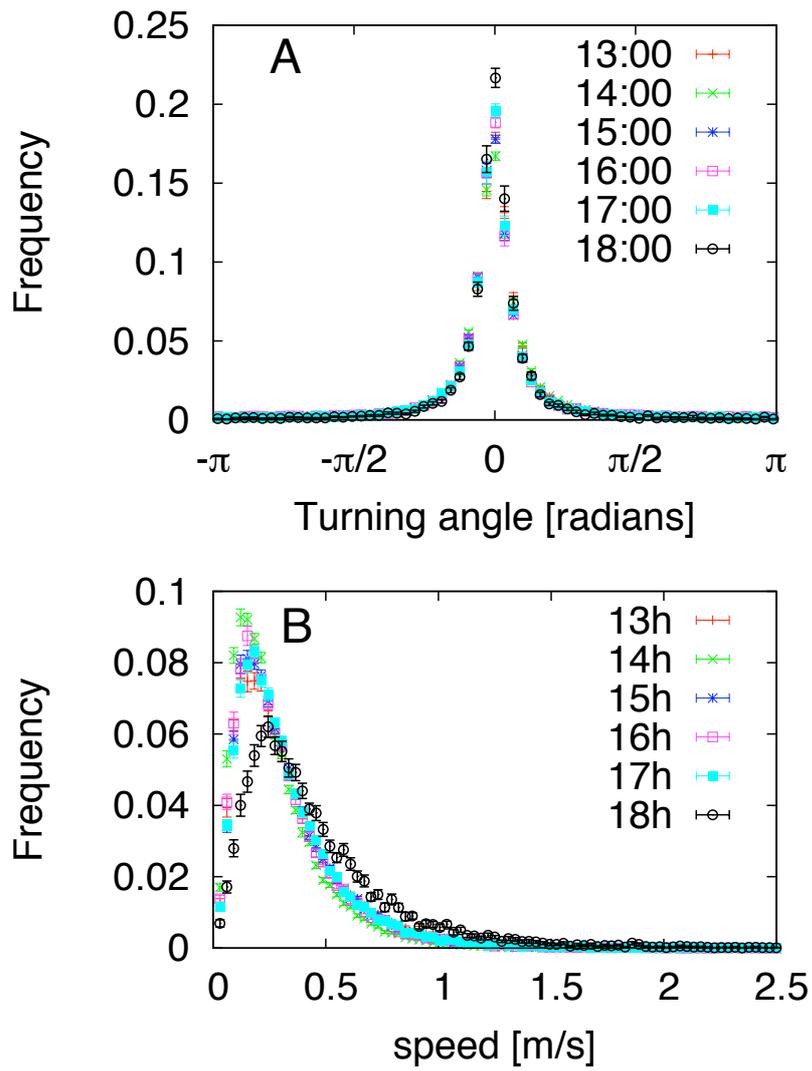

Figure 3: (A) Distribution of turning angles and (B) Distribution of the individual swimming speed at different hours.

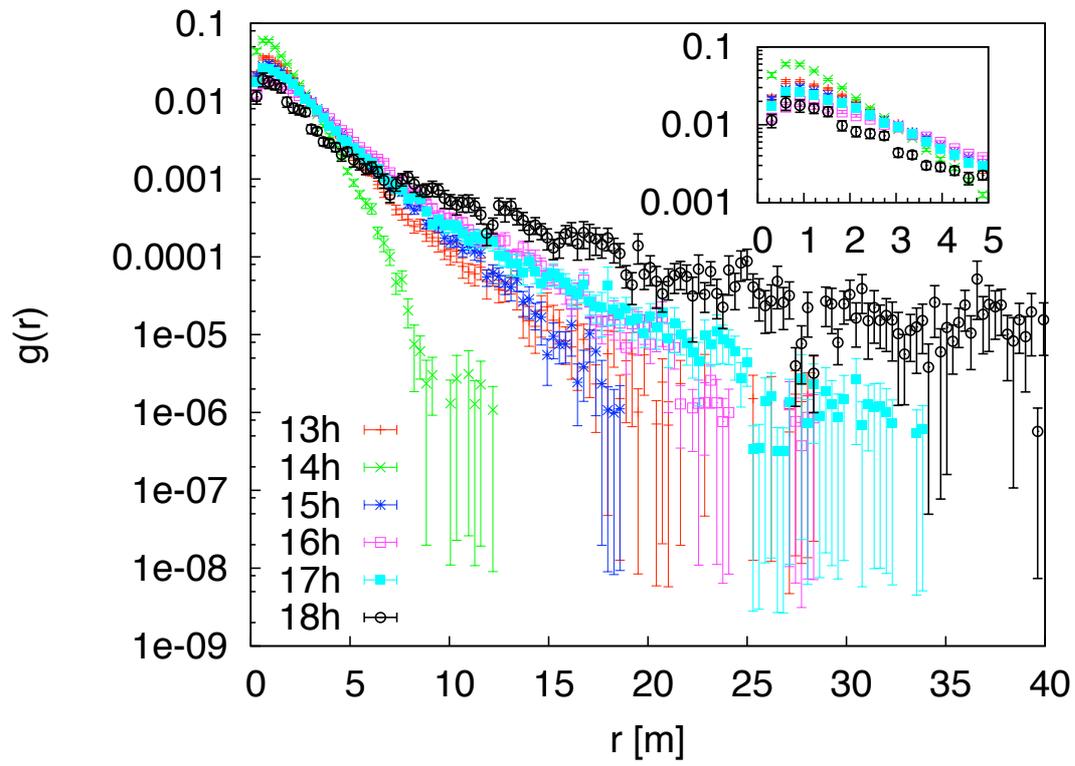

Figure 4: Fish pair-distribution function (Eq. 1) at different hours. Inset: zoom of g(r) at small r.

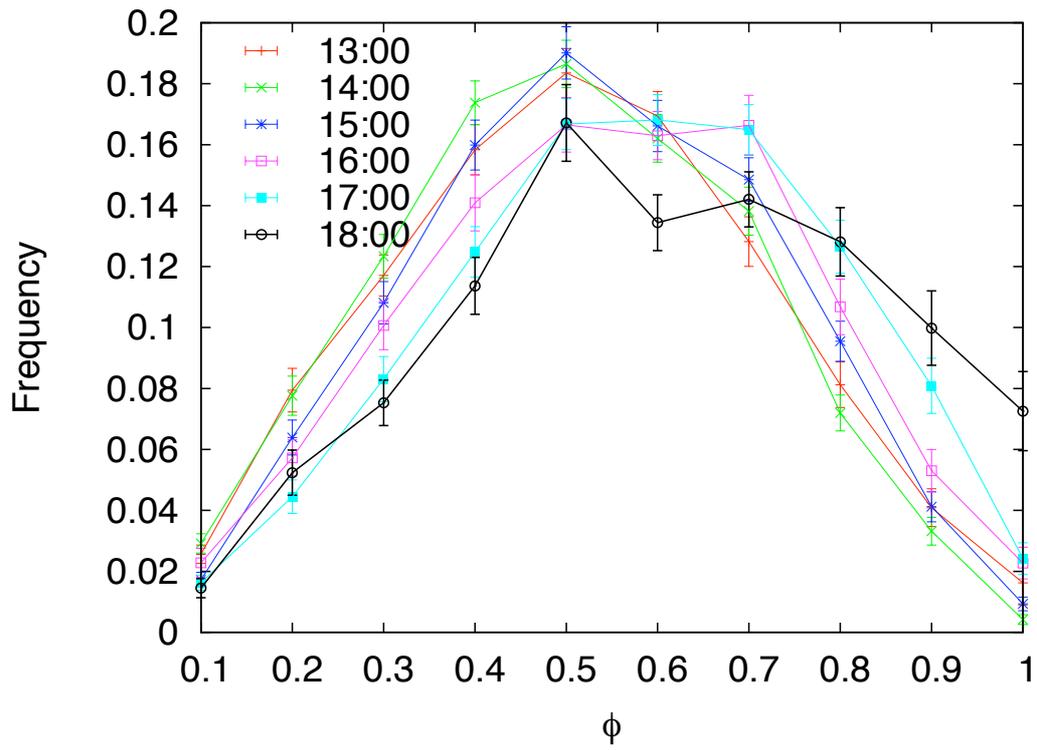

Figure 5: Distribution of the tagged-fish polarization (Eq. 2) at different hours.